\documentclass{PoS}
\pdfoutput=1
\usepackage{bm}

\newcommand{\beq}{\begin{equation}}
\newcommand{\eeq}{\end{equation}}
\newcommand{\beqy}{\begin{eqnarray}}
\newcommand{\eeqy}{\end{eqnarray}}

\newcommand{\me}[3]{\langle #1\vert\ #2\ \vert #3\rangle}
\newcommand{\ex}{\widehat{\bm{x}}}
\newcommand{\ey}{\widehat{\bm{y}}}
\newcommand{\ez}{\widehat{\bm{z}}}

\title{Excited isovector mesons using the stochastic LapH method}

\ShortTitle{Excited isovectors}

\author{\speaker{Colin Morningstar}$^a$, Brendan Fahy$^{a,b}$, 
        You-Cyuan Jhang$^a$, Keisuke~J.~Juge$^{c}$, David~Lenkner$^d$, and 
        Chik Him~Wong$^e$\\
\llap{$^a$} Dept.~of Physics, Carnegie Mellon University, 
        Pittsburgh, PA 15213, USA\\
\llap{$^b$} High Energy Accelerator Research Organization (KEK), 
        Ibaraki 305-0801, Japan\\
\llap{$^c$} Dept.~of Physics, University of the Pacific, 
        Stockton, CA 95211, USA\\
\llap{$^d$} Data Science Automation, 375 Valley Brook Road,
            Pittsburgh, PA 15317, USA\\
\llap{$^e$} Department of Physics, University of Wuppertal, 
            Gaussstrasse 20, D-42119, Germany}

\abstract{The spectrum of excited isovector mesons is studied using a 
$32^3 \times 256$ anisotropic lattice with $u,d$ quark masses set to give a pion mass 
near 240 MeV.  Results in the bosonic isovector nonstrange symmetry channels of zero 
total momentum are presented using correlation matrices of unprecedented size.  In addition 
to spatially-extended single-meson operators, large numbers of two-meson operators are used, 
involving a wide variety of light isovector, isoscalar, and strange meson operators of 
varying relative momenta.  All needed Wick contractions are efficiently evaluated using a 
stochastic method of treating the low-lying modes of quark propagation that exploits 
Laplacian Heaviside quark-field smearing.  Level identification is discussed.}

\FullConference{The 32nd International Symposium on Lattice Field Theory,\\
		23-28 June, 2014\\
		Columbia University New York, NY}

\begin{document}

\section{Introduction}

In a series of papers\cite{baryons2005A,baryon2007,nucleon2009,
Bulava:2010yg,StochasticLaph,ExtendedHadrons}, we have been striving to 
compute the finite-volume stationary-state energies of QCD using Markov-chain
Monte Carlo integration of the QCD path integrals formulated on a
space-time lattice.  In this talk, our progress made during the past year is 
described.  Last year, we presented our first results in the zero-momentum 
bosonic $I=1,\ S=0,\ T_{1u}^+$ symmetry sector of QCD on small $24^3\times 128$
lattices with an unphysically heavy pion mass around 390~MeV.  Here, we
report on results obtained on a $32^3\times 256$ anisotropic lattice for which
the pion mass is around 240~MeV. All needed Wick contractions are efficiently 
evaluated using a stochastic method of treating the low-lying modes of quark
propagation that exploits Laplacian Heaviside quark-field smearing.   Given 
the large number of levels extracted, level identification becomes a key issue.

\section{Operators, configurations, and analysis}
\label{sec:ops}

The stationary-state energies in a particular symmetry sector can be extracted 
from an $N\times N$ Hermitian correlation matrix 
   $ {\cal C}_{ij}(t)
   = \langle 0\vert\, O_i(t\!+\!t_0)\, \overline{O}_j(t_0)\ \vert 0\rangle,
   $
where the $N$ operators $\overline{O}_j$ act on the vacuum to create the states 
of interest at source time $t_0$ and are accompanied by conjugate operators $O_i$ 
that can annihilate these states at a later time $t+t_0$.  
Estimates of ${\cal C}_{ij}(t)$ are obtained with the Monte Carlo method
using the stochastic LapH method\cite{StochasticLaph} which allows all needed
quark-line diagrams to be computed.  

All of our single-hadron operators are assemblages of basic building blocks
which are gauge-covariantly-displaced, LapH-smeared quark fields, as described
in Refs.~\cite{baryons2005A,StochasticLaph,ExtendedHadrons}.  Each of our
single-hadron operators creates and annihilates a definite momentum.
Group-theoretical projections are used to construct operators that transform 
according to the irreducible representations of the space group $O_h^1$, 
plus $G$-parity, when appropriate.  In order to build up the necessary orbital
and radial structures expected in the hadron excitations, we use a variety of
spatially-extended configurations.
For practical reasons, we restrict our attention to certain classes of 
momentum directions for the single hadron operators: on axis 
$\pm\ex,\ \pm\ey,\ \pm\ez$, planar diagonal 
$\pm\ex\pm\ey,\ \pm\ex\pm\ez,\ \pm\ey\pm\ez$,
and cubic diagonal $\pm\ex\pm\ey\pm\ez$.  However, some special momentum
directions, such as $\pm 2\ex\pm\ey$, are used.
We construct our two-hadron operators as superpositions of single-hadron 
operators of definite momenta.  Again, group-theoretical projections are 
employed to produce two-hadron operators that transform irreducibly under
the symmetry operations of our system.
This approach is efficient for creating large numbers of two-hadron 
operators, and generalizes to three or more hadrons.

In finite volume, all energies are discrete so that each correlator matrix
element has a spectral representation of the form
\beq
   {\cal C}_{ij}(t) = \sum_n Z_i^{(n)} Z_j^{(n)\ast}\ e^{-E_n t},
   \qquad\quad Z_j^{(n)}=  \me{0}{O_j}{n},
\eeq
assuming temporal wrap-around (thermal) effects are negligible.
We extract energies from our correlation matrices using a ``single rotation'' 
or ``fixed coefficient'' method.  Starting with a raw correlation matrix 
${\cal C}(t)$, we first try to remove the effects of differing normalizations 
by forming the matrix 
$ C_{ij}(t)={\cal C}_{ij}(t)\ (\ {\cal C}_{ii}(\tau_N){\cal C}_{jj}(\tau_N)\ )^{-1/2}$,
taking $\tau_N$ at a very early time, such as $\tau_N=3$.  We ensure that
$C$ is positive definite and has a reasonable condition number.  Standard
projection methods can be used to remove problematic modes.  We then
solve the generalized eigenvector problem $Ax=\lambda Bx$ with
$A=C(\tau_D)$ and $B=C(\tau_0)$ for particular choices of times $\tau_0$ and
$\tau_D$ (see below).   The eigenvectors 
obtained are used to ``rotate'' the correlator $C(t)$ into a correlator 
$G(t)$ for which $G(\tau_0)=1$, the identity matrix, 
and $G(\tau_D)$ is diagonal.  At other times, $G(t)$ 
need not be diagonal.  However, with judicious choices of $\tau_0$ and $\tau_D$, 
one finds that the off-diagonal elements of $G(t)$ remain zero within 
statistical precision 
for $t>\tau_D$. The rotated correlator is given by
\beq
 G(t) = U^\dagger\ C(\tau_0)^{-1/2}\ C(t)\ C(\tau_0)^{-1/2}\ U,
\label{eq:rotatedcorr}
\eeq
where the columns of $U$ are the orthonormalized eigenvectors of
$C(\tau_0)^{-1/2}\ C(\tau_D)\ C(\tau_0)^{-1/2}$.
Rotated effective masses can then be defined by
\beq
  m_G^{(n)}(t)=\frac{1}{\Delta t}
  \ln\left(\frac{G_{nn}(t)}{
  G_{nn}(t+\Delta t)}\right),
\label{eq:roteffmass}
\eeq
which tend to the lowest-lying $N$ stationary-state energies
produced by the $N$ operators.  Correlated-$\chi^2$ fits to 
the estimates of $G_{nn}(t)$ using the forms 
\beq
A_n e^{-E_n\,t} \left(1+B_n e^{-\Delta_n^2\,t}\right)
+ A_n e^{-E_n\,(T-t)}\left(1+B_n e^{-\Delta_n^2\,(T-t)}\right),
\eeq
where $T$ is the temporal extent of the lattice, yield the energies $E_n$
and the overlaps $A_n$ to the rotated operators for each $n$. Using the 
rotation coefficients, one can then easily obtain the overlaps 
$Z^{(n)}_j=C(\tau_0)^{1/2}_{jk}\ U_{kn}\ A_n$ (no summation over $n$)
corresponding to the rows and columns of the correlation matrix $C(t)$.

We are currently focusing on three Monte Carlo ensembles: (A) a set of 
412 gauge-field configurations on a large $32^3\times 256$ anisotropic lattice 
with a pion mass $m_\pi\sim 240$~MeV, (B) an ensemble of 551 configurations
on an $24^3\times 128$ anisotropic lattice with a pion mass
$m_\pi\sim 390$~MeV, and (C) an ensemble of 584 configurations
on an $24^3\times 128$ anisotropic lattice with a pion mass
$m_\pi\sim 240$~MeV.  We refer to these ensembles as the 
$(32^3\vert 240)$, $(24^3\vert 390)$, and $(24^3\vert 240)$ ensembles,
respectively.  These ensembles were generated using the Rational 
Hybrid Monte Carlo (RHMC) algorithm\cite{Clark:2004cp}. In each ensemble, successive 
configurations are separated by 20 RHMC trajectories to minimize autocorrelations.
An improved anisotropic clover fermion action and an improved gauge field 
action are used\cite{Lin:2008pr}.  In these ensembles, $\beta=1.5$
and the $s$ quark mass parameter is set to $m_s=-0.0743$ in order to reproduce 
a specific combination of hadron masses\cite{Lin:2008pr}.  
In the $(24^3\vert 390)$ ensemble, the light quark mass parameters are set to
$m_u=m_d=-0.0840$ so that the pion mass is around 390~MeV if one sets the scale 
using the $\Omega$ baryon mass.   In the $(32^3\vert 240)$ and 
$(24^3\vert 240)$ ensembles, $m_u=m_d=-0.0860$ are used, resulting in a pion 
mass around 240~MeV.  The spatial grid size is $a_s\sim 0.12$~fm, whereas
the temporal spacing is $a_t\sim 0.035$~fm.

In our operators, a stout-link staple weight $\xi=0.10$ is used with
$n_\xi=10$ iterations.  For the cutoff in the LapH smearing, we use 
$\sigma_s^2=0.33$,  which translates into the number $N_v$ of LapH eigenvectors 
retained being $N_v=112$ for the $24^3$ lattices and $N_v=264$ for the $32^3$ 
lattice. We use $Z_4$ noise in all of our stochastic estimates of quark propagation.
Our variance reduction procedure is described in Ref.~\cite{StochasticLaph}.
On the $24^3$ lattices, we use 4 widely-separated source times $t_0$,
and 8 are used on the $32^3$ lattice.

\section{Energies in the $T_{1u}^+$ channel} 
\label{sec:results}

\begin{figure}[t]
  \centering
  \includegraphics[width=5.0in]{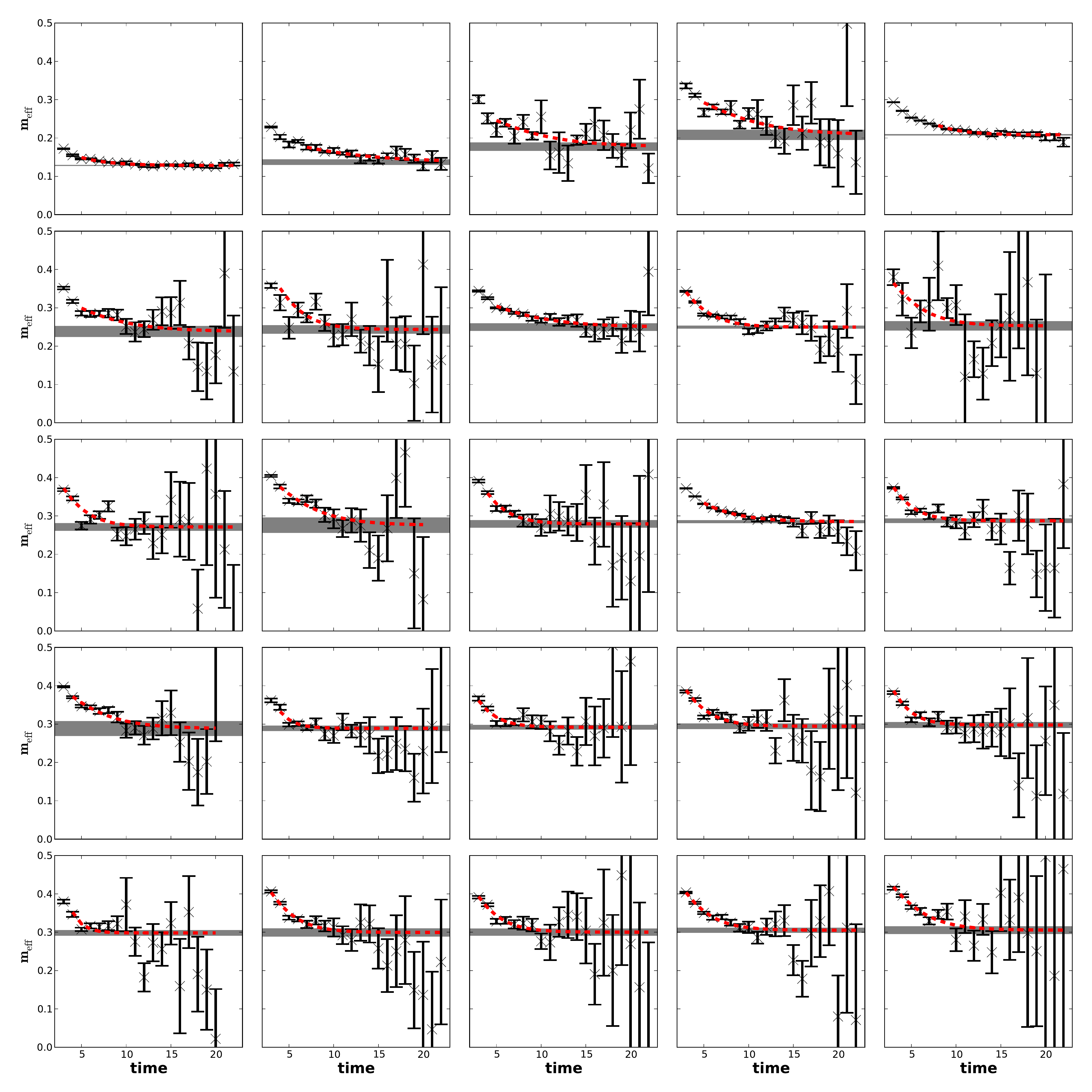}
  \caption[First 25 effective masses for $T_{1u}^+$]{
  Rotated effective masses $m_G^{(n)}(t)$ 
  (see Eq.~(\ref{eq:roteffmass})) for the 25 lowest-lying energy levels in the 
  zero-momentum bosonic $I=1,\ S=0,\ T_{1u}^+$ channel for the $(32^3\vert 240)$ 
  ensemble using 14 single-meson operators, 23 isovector+isovector operators, 
  31 light-isoscalar+isovector operators, 31 $\overline{s}s$-isoscalar+isovector
  operators, and 9 kaon+antikaon operators.  Dashed lines indicate
  energy extractions from correlated-$\chi^2$ fits.  Gray bands show the best
  fit values of the energies, whose standard deviations are indicated by the
  width of each band.}
  \label{fig:emasses}
\end{figure}

We focus here on the resonance-rich $I=1,\ S=0,\ T_{1u}^+$ channel of total
zero momentum.  This channel has odd parity, even $G$-parity, and contains 
the spin-1 and spin-3 mesons.
Low statistics runs on smaller lattices led us to
include 14 particular single-meson (quark-antiquark) operators.
We took special care to include operators that could produce the 
spin-$3$ $\rho_3(1690)$ state, in addition to the other spin-$1$ states.
Low statistics runs also gave us the masses of the lowest-lying
mesons, such as the $\pi,\eta, K,$ and so on.  Given these known
mesons, we used software written in \textsc{Maple} to find all possible
two-meson states in our cubic box in this $T_{1u}^+$ symmetry
channel, assuming no energy shifts from interactions or the
finite volume.  We used these so-called ``expected two-meson levels''
to guide our choice of two-meson operators to include.  We
included 23 isovector-isovector meson operators, 31 operators
that combine an isovector with a light isoscalar (using only $u,d$
quarks), 31 operators that combine an isovector with an
$\overline{s}s$ isoscalar meson, and 9 kaon-antikaon operators.

We obtained results for the lowest 50 energy levels using the $(32^3\vert 240)$
ensemble from our $108\times 108$ correlation matrix.  The rotated effective 
masses $m_G^{(n)}(t)$  (see Eq.~(\ref{eq:roteffmass})) using $\tau_0=5$ and
$\tau_D=8$ are shown for the first 25 levels in Fig.~\ref{fig:emasses}.
The results shown here are not finalized yet.  We are still
varying the fitting ranges to improve the $\chi^2$, as needed in
some instances.  We are investigating the effects of adding more
operators, and we are even still verifying our analysis/fitting
software.  However, these figures do demonstrate that the extraction
of a large number of energy levels is indeed possible, and the
plots indicate the level of precision that can be attained with
our stochastic LapH method.  Keep in mind that we have not included 
any three-meson operators in our correlation matrix. 

With such a large number of energies extracted, level identification 
becomes a key issue.  QCD is a complicated interacting quantum field
theory, so characterizing its stationary states in finite volume
is not likely to be done in a simple way.  Level identification must
be inferred from the $Z$ overlaps of our probe operators, analogous
to deducing resonance properties from scattering cross sections
in experiments.  Judiciously chosen probe operators, constructed from
smeared fields, should excite the low-lying states of interest, with
hopefully little coupling to unwanted higher-lying states, and help
with classifying the levels extracted.  Small-$a$ classical
expansions can help to characterize the probe operators, and hence,
the states they produce.

\begin{figure}[t]
  \begin{center}
    \includegraphics[width=1.9in]{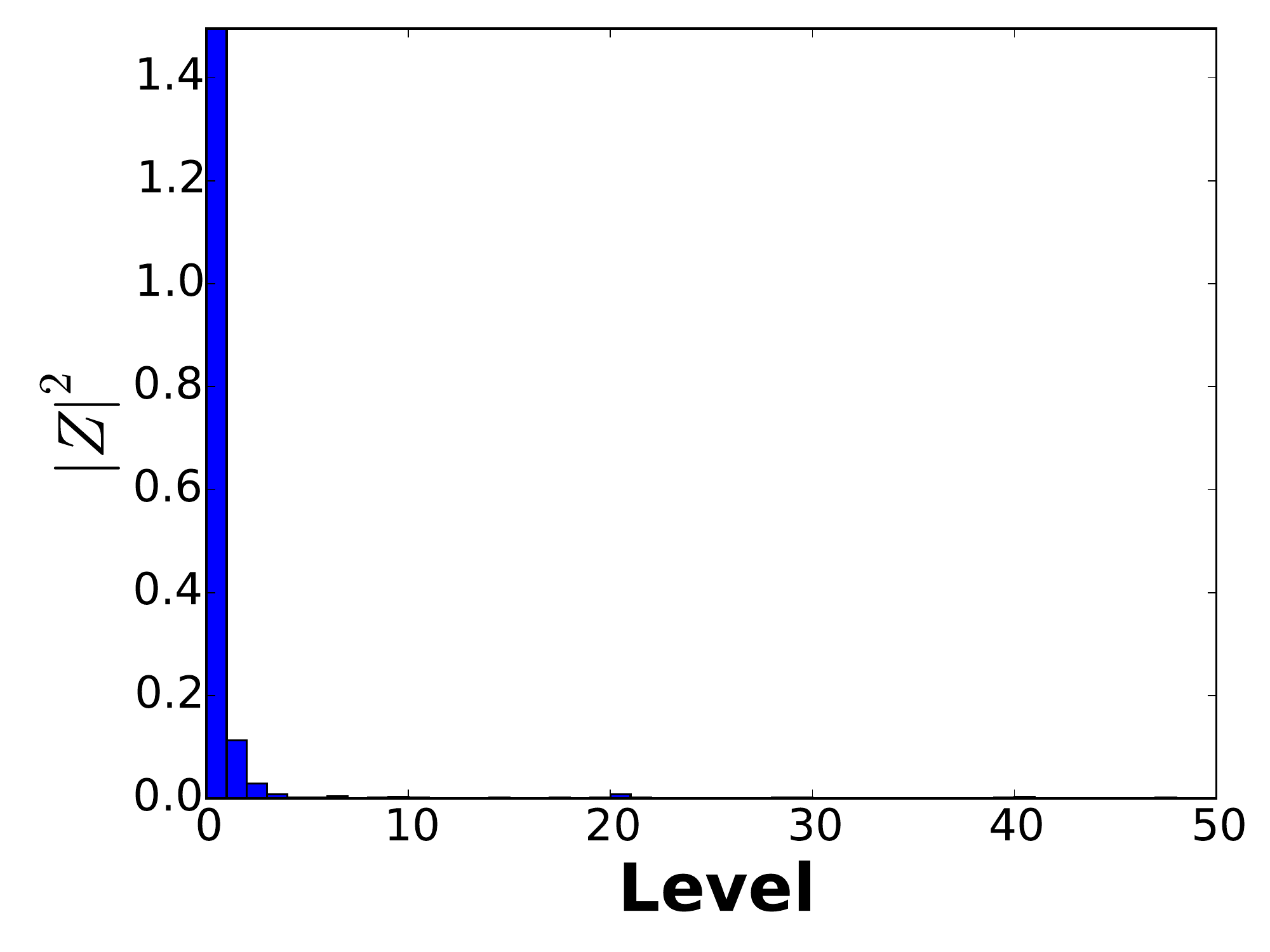}
    \includegraphics[width=1.9in]{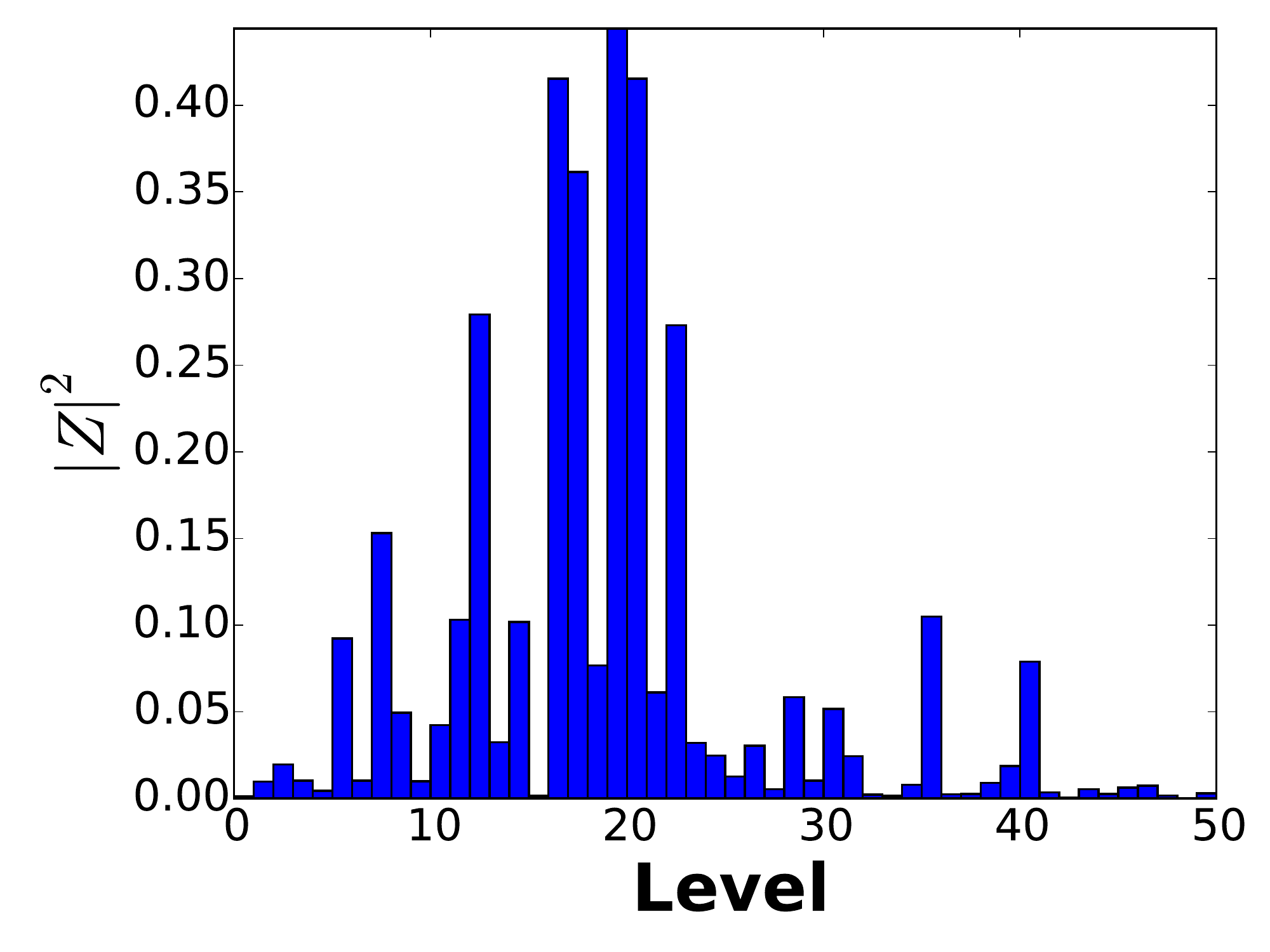}
    \includegraphics[width=1.9in]{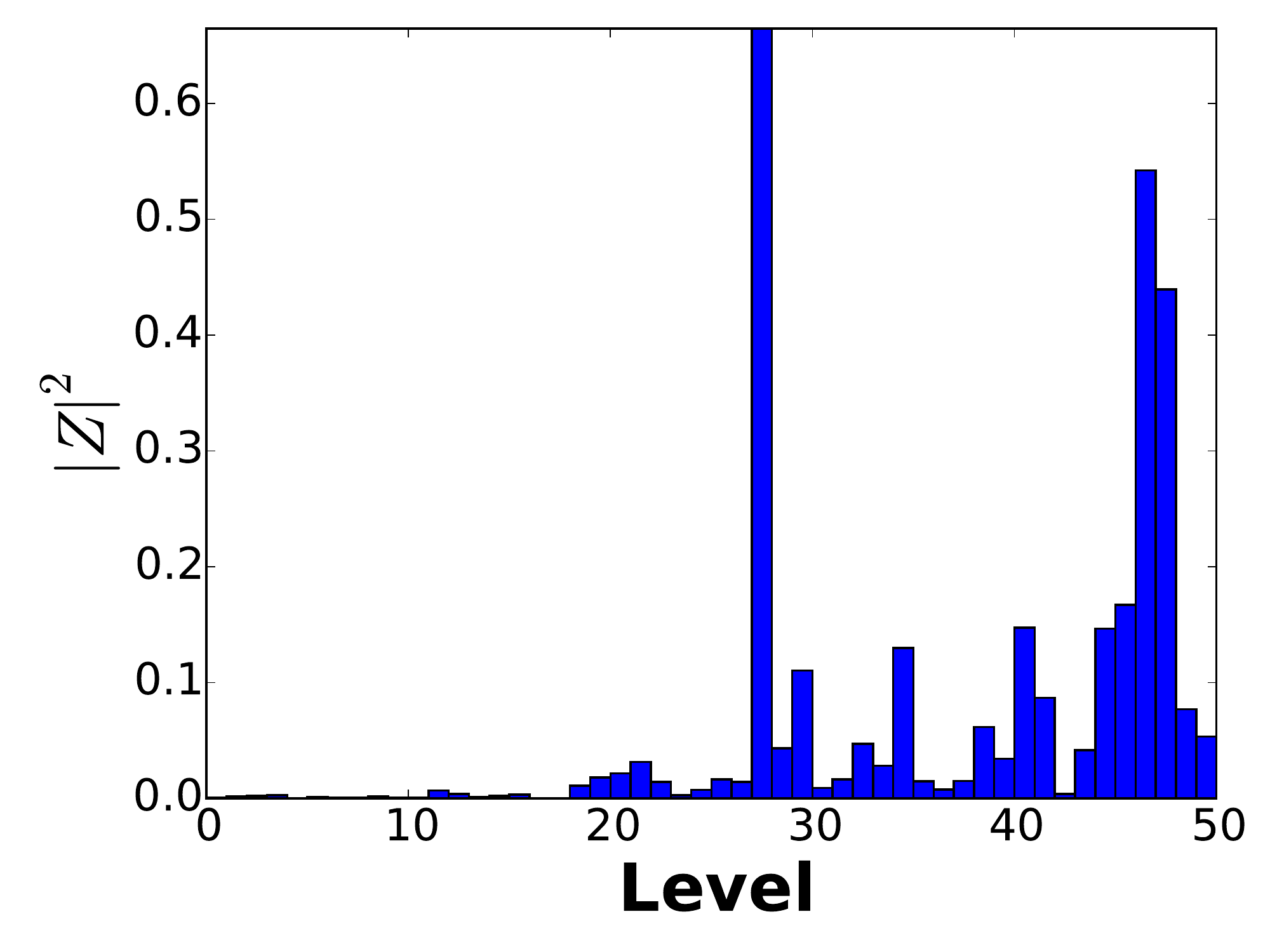}\\
    \includegraphics[width=1.9in]{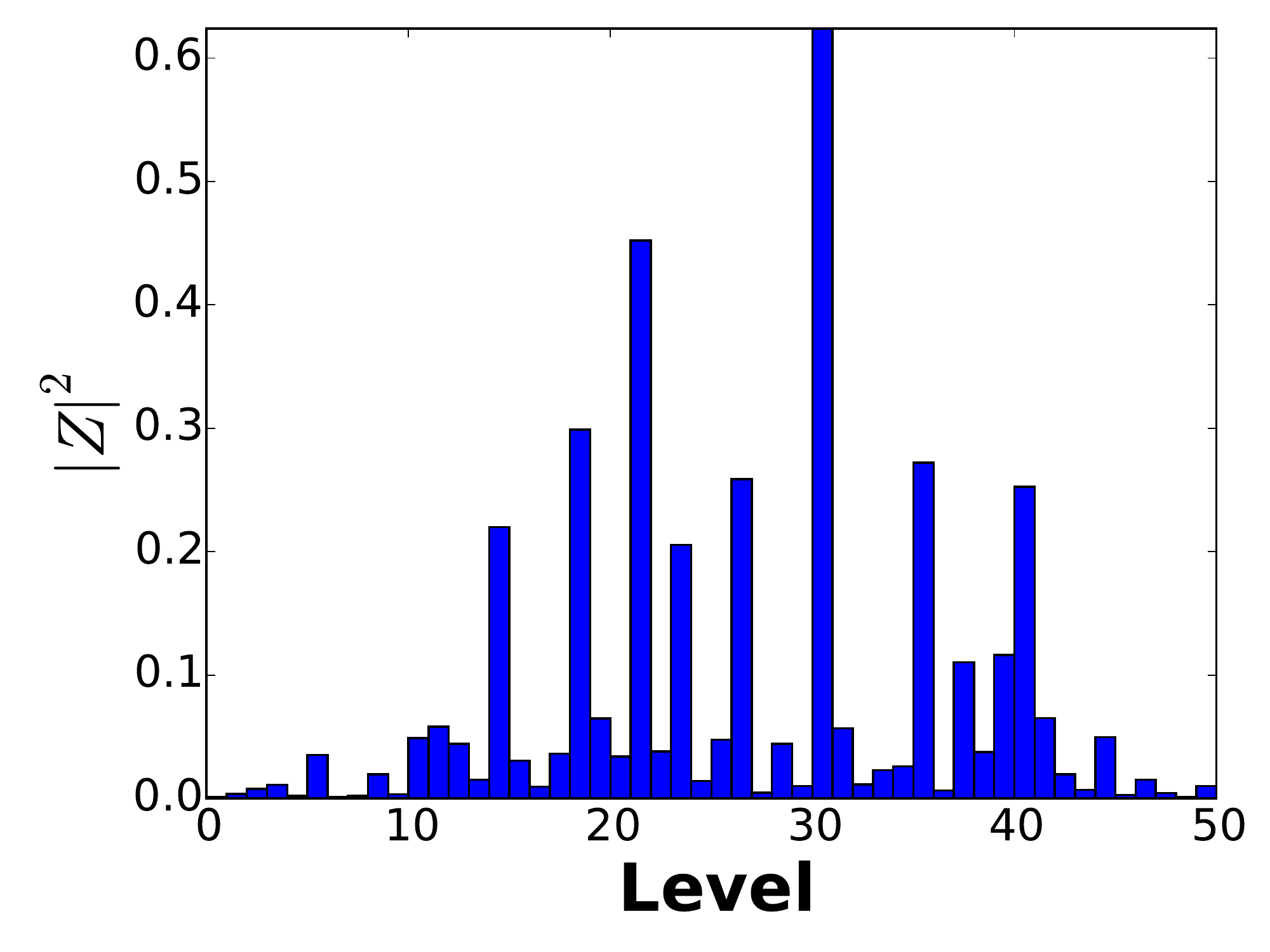}
    \includegraphics[width=1.9in]{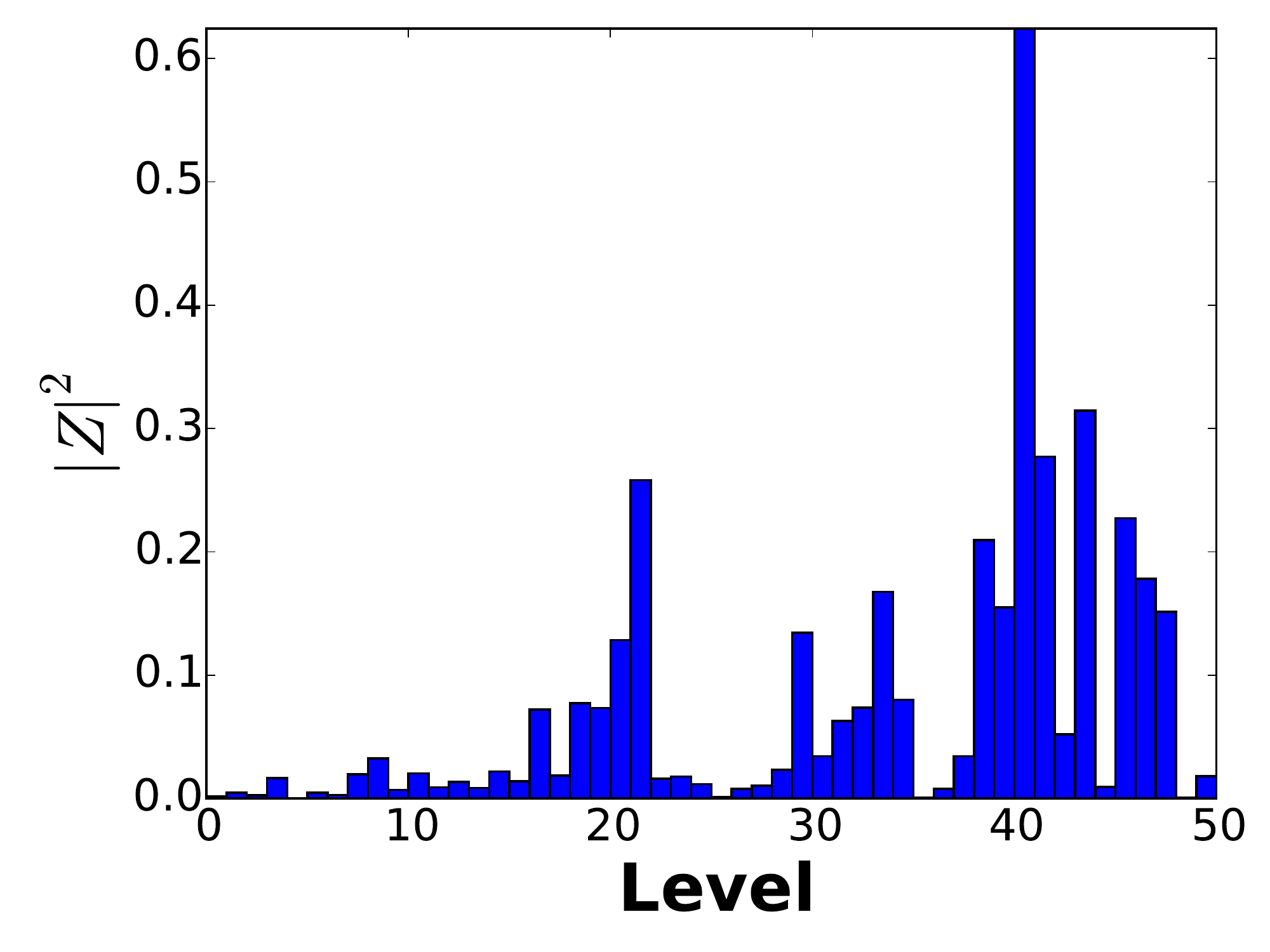}
    \includegraphics[width=1.9in]{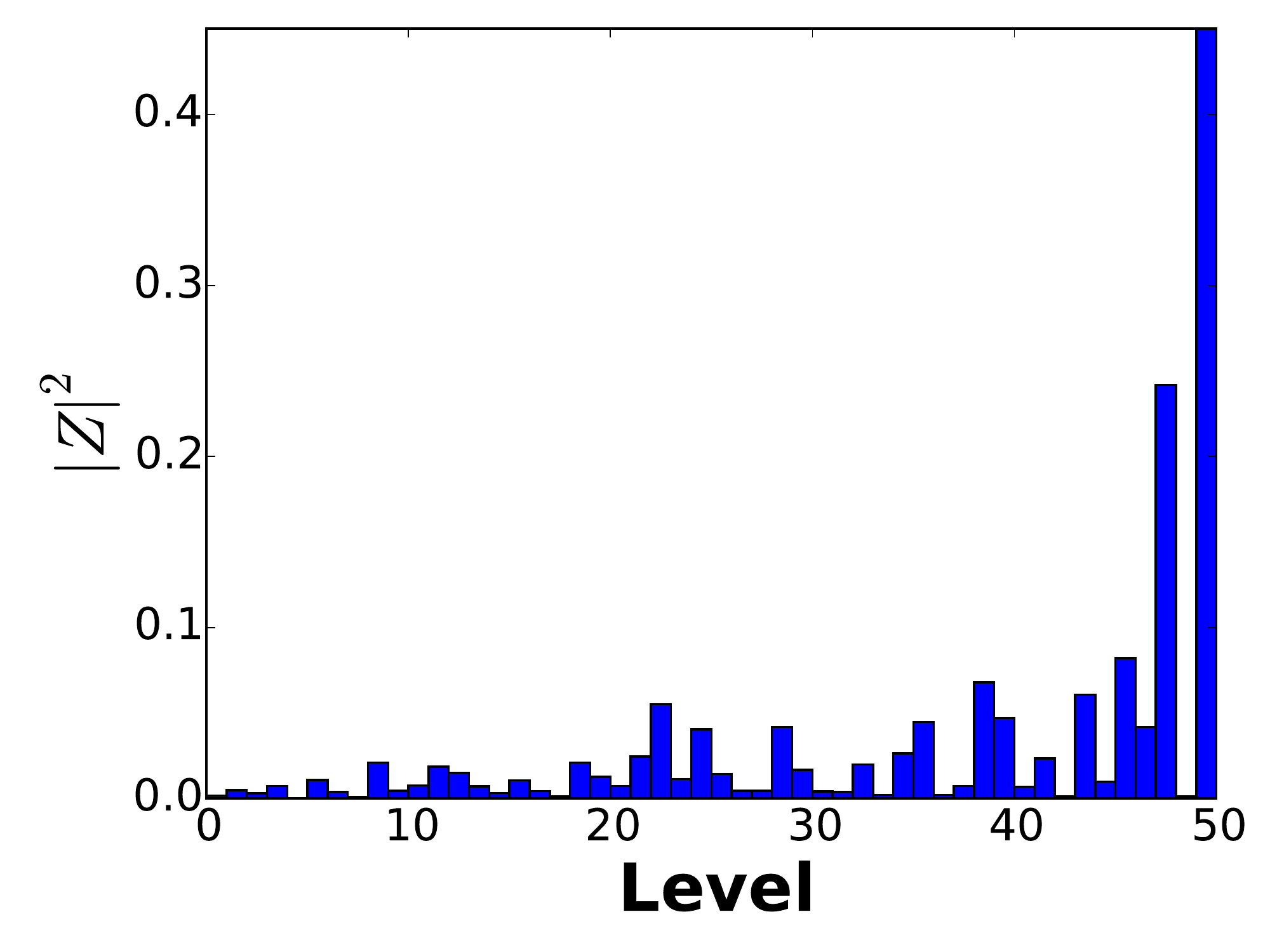}
\end{center}
\caption[Optimized single-hadron operator overlaps]{
Overlaps $\vert \widetilde{Z}^{(n)}_j\vert^2$ of ``optimized'' single-hadron
operator $\widetilde{O}_j$ against the eigenstates labelled by $n$.
The overall normalization is arbitrary in each plot.
\label{fig:shopt}}
\end{figure}

We particularly wish to identify the finite-volume stationary-state levels 
expected to evolve into the single-meson resonances corresponding to 
quark-antiquark excitations in infinite volume. To accomplish this, we utilize
``optimized'' single-hadron operators as our probes.  We first restrict
our attention to the $14\times 14$ correlator matrix involving only
the 14 chosen single-hadron operators.  We then perform an optimization
rotation to produce so-called ``optimized'' single-hadron (SH) operators
$\widetilde{O}_j$, which are linear combinations of the 14 original
operators, determined in a manner analogous to
Eq.~(\ref{eq:rotatedcorr}).  We order these SH-optimized operators according
to their effective mass plateau values, then evaluate the overlaps 
$\widetilde{Z}_j^{(n)}$ for these SH-optimized operators using
our analysis of the full $108\times 108$ correlator matrix.  The
results are shown in Fig.~\ref{fig:shopt}.

\begin{figure}[t]
  \begin{center}
  \includegraphics[width=5.0in]{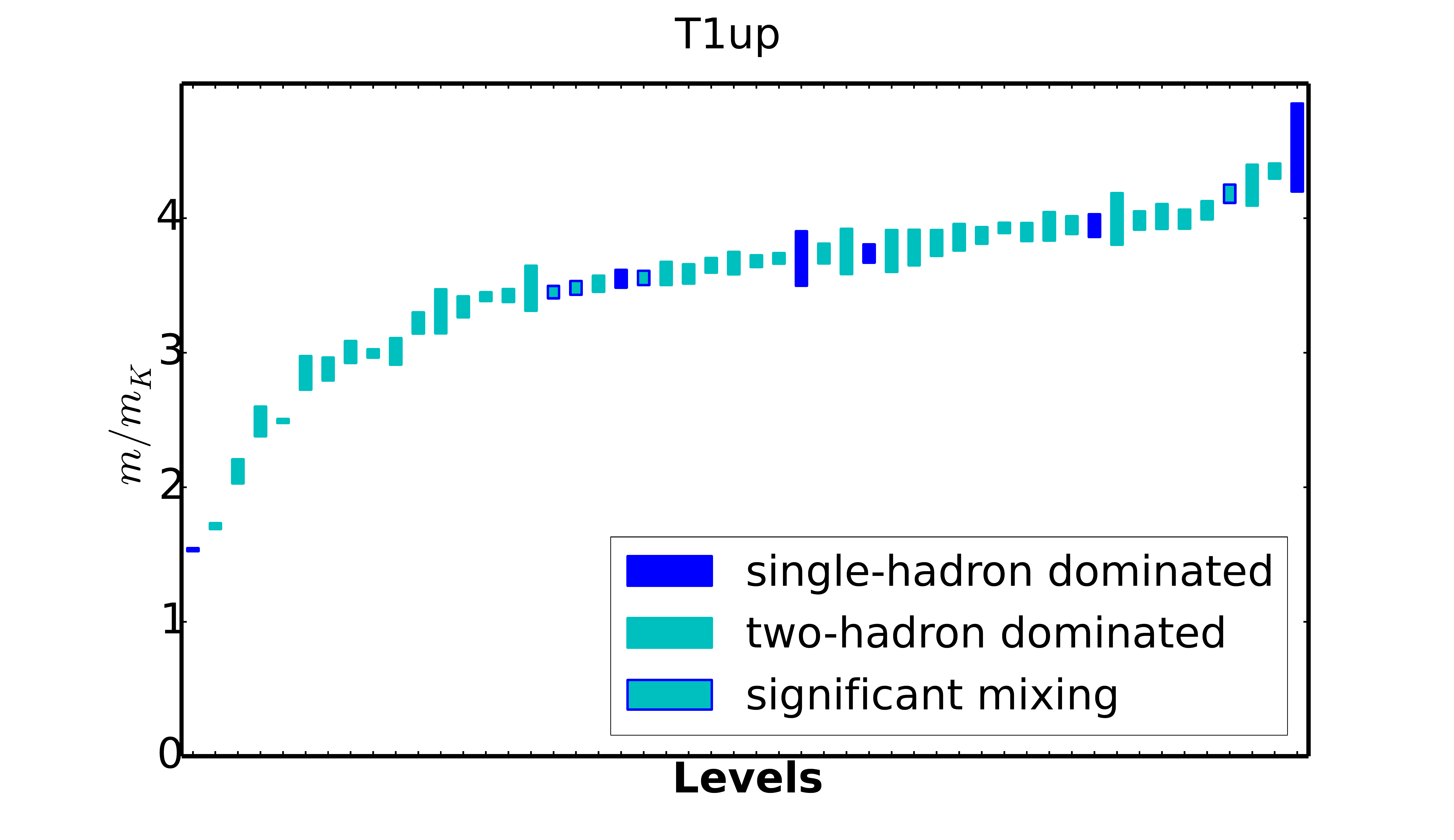}
  \end{center}
  \vspace*{-8mm}
  \caption[Spectrum of the lowest 50 levels in $T_{1u}^+$]{
    Energies $m$ as ratios of the kaon mass $m_K$ for the first fifty states 
    excited by our single- and two-hadron operators in the
    $T_{1u}^+$ channel.  For each optimized single-hadron operator,
    the level of maximum overlap is indicated by a solid blue box, and
    levels with overlaps greater than $75\%$ of the largest are indicated 
    by a dark blue outline.}
  \label{fig:staircase}
\end{figure}

\begin{figure}[tb]
  \begin{center}
  \includegraphics[width=4.0in]{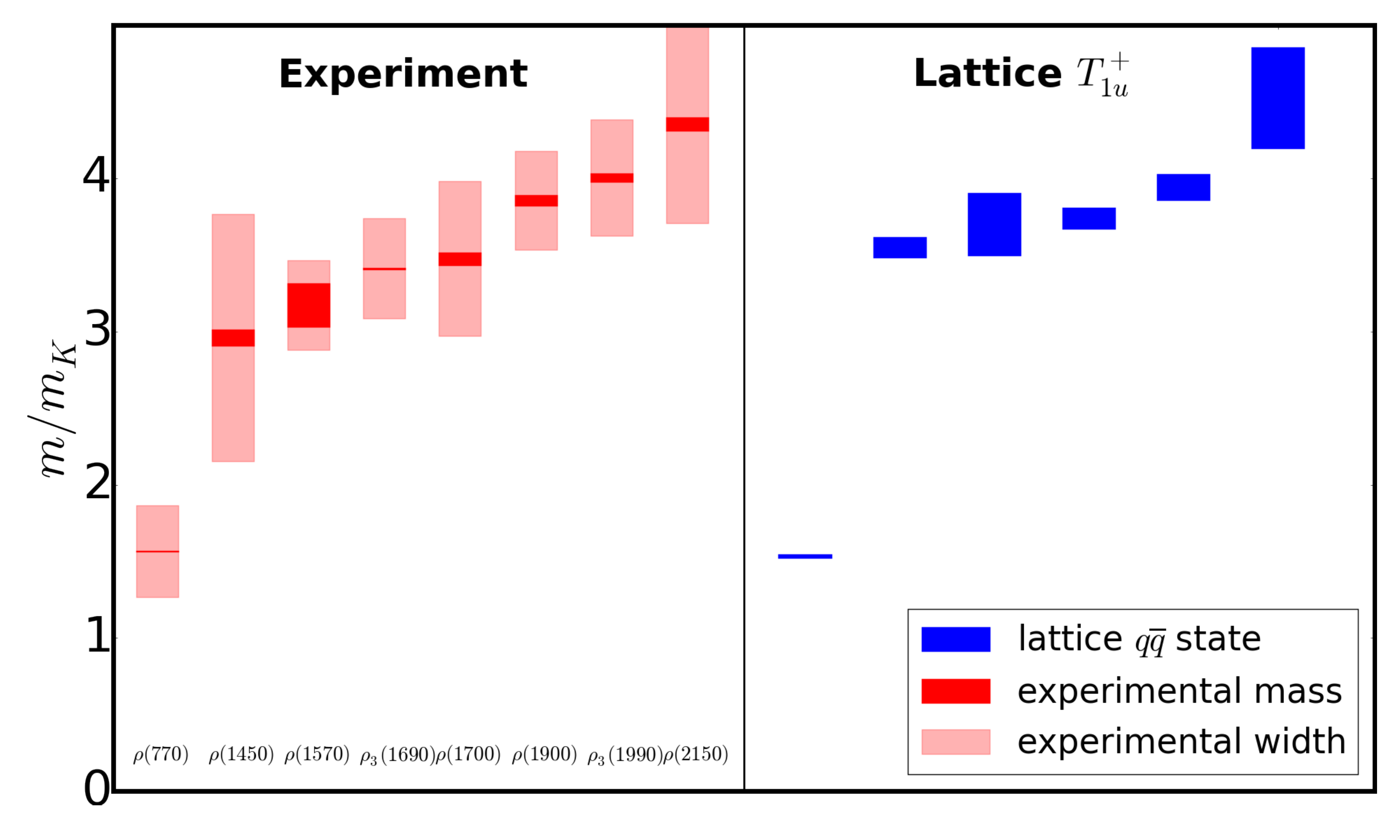}
  \end{center}
  \vspace*{-7mm}
  \caption[$T_{1u}^+$ spectrum compared to experiment.]{
    Comparison of
    the experimental spectrum of resonances with our finite-volume
    energies corresponding to quark-antiquark excitations. All masses $m$ are
    shown as ratios over the kaon mass $m_K$.  In the left
    hand side, dark red boxes indicate the experimental
    masses, with the vertical heights showing the uncertainties in the mass
    measurements.  The light red boxes indicate the experimental widths
    of the resonances.  In the right hand side, our masses for the
    quark-antiquark excitations are shown
    by dark blue boxes, whose heights indicate statistical uncertainties
    only. This $T_{1u}^+$ channel includes
    both $\rho$ (spin 1) and $\rho_3$ (spin 3) states. }
  \label{fig:experiment}
\end{figure}

Our energies in the $T_{1u}^+$ channel are summarized by the ``staircase'' plot
in Fig.~\ref{fig:staircase}.  For each SH optimized operator, the level
with the largest overlap is identified on this plot using a solid blue box.
Other levels with significant overlaps with the SH optimized operator are
indicated by boxes with a dark blue outline.  The remaining cyan boxes are
levels with overlaps dominated by two-meson operators.  The energies of the
levels with solid blue boxes are collected and shown in Fig.~\ref{fig:experiment},
which compares these energies to experiment.  The finite-volume energies
should agree with experiment only within the widths of the infinite-volume
resonances.  We believe we have extracted all meson resonances that
are quark-antiquark excitations.  One observes more levels in experiment,
although the experimental observations are controversial in some cases.
Keep in mind that resonances that are not quark-antiquark excitations, such as 
so-called molecular states, would not be identified by our SH optimized 
operator overlaps.   Again, we mention that three and four meson states 
are not taken into account at all.

\section{Conclusion}
\label{sec:conclude}

In this talk, our progress made during the past year in computing
the finite-volume stationary-state energies of QCD was described.
Our results in the zero-momentum bosonic $I=1,\ S=0,\ T_{1u}^+$ symmetry 
sector of QCD for the $(32^3\vert 240)$ ensemble using a correlation matrix
of 108 operators were presented.  All needed Wick contractions were 
efficiently evaluated using  the stochastic LapH method.   Issues related 
to level identification were discussed.

This work was supported by the U.S.~NSF under awards PHY-1306805 and PHY-1318220, 
and through TeraGrid/XSEDE resources provided by 
TACC and NICS under grant number TG-MCA07S017.

\end{document}